\documentclass{emulateapj}
\usepackage{graphicx}
\usepackage{amsmath, amssymb, amsfonts}
\usepackage{epstopdf}
\usepackage{natbib}
\shortauthors{Paterno-Mahler et al.}
\shorttitle{The Sloshing Spiral in A2029}
\submitted{Accepted by the Astrophysical Journal}
\begin{document}

\title{Deep \textit{Chandra} Observations of the Extended Gas Sloshing Spiral in A2029}
\author{R. Paterno-Mahler\altaffilmark{1}, E.L. Blanton\altaffilmark{1}, S.W. Randall\altaffilmark{2},  and T.E. Clarke\altaffilmark{3}}
\altaffiltext{1}{Astronomy Department and Institute for Astrophysical Research, Boston University, 725 Commonwealth Ave, Boston, MA 02215, USA; rachelpm@bu.edu, eblanton@bu.edu}
\altaffiltext{2}{Harvard Smithsonian Center for Astrophysics, 60 Garden Street, Cambridge, MA 02138, USA; srandall@head.cfa.harvard.edu}
\altaffiltext{3}{Naval Research Laboratory, Code 7213, 4555 Overlook Avenue Southwest, Washington DC, 20375, USA; tracy.clarke@nrl.navy.mil}

\begin{abstract} 
Recent X-ray observations of galaxy clusters have shown that there is substructure present in the intracluster medium (ICM), even in clusters that are seemingly relaxed.  This substructure is sometimes a result of sloshing of the ICM, which occurs in cool core clusters that have been disturbed by an off-axis merger with a sub-cluster or group.  We present deep \textit{Chandra} observations of the cool core cluster Abell 2029, which has a sloshing spiral extending radially outward from the center of the cluster to approximately 400~kpc at its fullest extent---the largest continuous spiral observed to date.  We find a surface brightness excess, a temperature decrement, a density enhancement, an elemental abundance enhancement, and a smooth pressure profile in the area of the spiral.  The sloshing gas seems to be interacting with the southern lobe of the central radio galaxy, causing it to bend and giving the radio source a wide-angle tail (WAT) morphology.  This shows that WATs can be produced in clusters that are relatively relaxed on large scales. We explore the interaction between heating and cooling in the central region of the cluster.  Energy injection from the active galactic nucleus (AGN) is likely insufficient to offset the cooling, and sloshing may be an important additional mechanism in preventing large amounts of gas from cooling to very low temperatures.
\end{abstract}


\section{Introduction}
Galaxy clusters are filled with an intracluster medium (ICM) of hot, X-ray emitting gas.  X-ray observations of this gas have shown that it can have significant substructure, even in seemingly relaxed clusters.  This substructure is sometimes related to the merger history of the cluster.   

Some of the most obvious features are cold fronts, which appear as surface brightness discontinuities across which the density and temperature sharply change while the pressure does not~\citep{review}.  Spiral features, which are directly related to cold fronts, have been observed in clusters such as Perseus~\citep{perseus, fabian2006}, Virgo~\citep{virgo}, and A496~\citep{a496}, as well as many others~\citep{lagana}.  The radial extents of these spirals range from less than 50~kpc for the Centaurus cluster to approximately 245~kpc for A85.  The spiral in A496 extends out to approximately 150~kpc~\citep{a496}, while the Virgo cluster shows a cold front feature out to 90~kpc~\citep{virgo}.  The spiral in the Perseus cluster extends to about 150~kpc~\citep{perseus}.  \citet{simionescu} found evidence that the spiral in Perseus may extend to 700~kpc in the east and over a Mpc in the west; however those cold fronts do not connect to form a continuous spiral despite the very deep available observations.

Cold fronts and spirals are created by mergers, which displace the cool, dense cluster gas from the center of the potential well.  These features have also been seen in simulations~\citep{am, zuhone, zuhone2012}, which show that a spiral can be created by an off-center minor merger with a subcluster.  These simulations are idealized binary mergers.  \citet{am} show explicitly that sloshing gas can create cold fronts and they are a consequence of minor mergers, while the \citet{zuhone, zuhone2012} simulations explore the effects of magnetized gas and viscosity on the properties of the spiral.  \citet{roediger} and \citet{a496} perform simulations of Virgo and A496 to directly compare simulations to observations.  The simulations of A496 explore sloshing in an elliptical cluster, which is pertinent to A2029.  Since the initial fly-by in these simulations is off-center, the cool gas in the cluster acquires angular momentum and does not fall back into the cluster center radially.  The cold fronts created by this sloshing are not concentric and combine to form a spiral feature.  These features can persist for billions of years~\citep{am}, and are thus not necessarily related to recent mergers.  The substructure can provide clues to the merger history of the cluster, providing insight into cluster formation.  Minor merger events such as the ones that cause cold fronts are estimated to occur at a rate of $\sim1/3$~event per halo per Gyr~\citep{ghizzardi}. 

Spiral morphology is an unambiguous indicator of gas sloshing.  Cold fronts can be caused by mergers or sloshing, which can lead to confusion as to the origin of the observed features, as in Abell 3667~\citep{3667}.  The orientation of the subcluster orbit can be constrained by knowing the orientation of the full extent of the spiral and the orientation of the brightness asymmetries induced by the sloshing~\citep{a496}.  On large scales, extended spirals provide a mechanism for transporting metals from the center of the cluster to the outskirts.  This has been seen observationally in Virgo~\citep{virgo} and A2052~\citep{blanton}, and in simulations such as \citet{roediger}, although other simulations do not show signs of global broadening in the abundance distribution~\citep{a496}.

Sloshing features could also provide part of the solution for the ``cooling flow" problem.  In a ``cooling flow" scenario, the central region of a galaxy cluster begins to lose energy via radiation.  The cooling gas contracts to maintain its pressure, which in turn causes the gas to cool even faster.  In order to maintain hydrostatic equilibrium, gas continues to flow in \citep[for a review, see][]{cf}. With such large amounts of gas being deposited in the center of the cluster, one would expect to see significant star formation, absent a significant heating mechanism.  X-ray observations of such clusters show that while there is some gas cooling to low temperatures and star formation is seen, the majority of the gas only drops to approximately $1/2$ to $1/3$ of the ambient temperature~\citep{peterson}, rather than cooling all the way to very low temperatures, and thus a heating mechanism is necessary.  Clusters with strong cooling flows ($t_{cool}< 1$~Gyr) are likely regulated by a feedback loop involving active galactic nuclei (AGN)~\citep{mcnamara}.  Secondary processes, such as sloshing, can also help counteract radiative losses when not enough energy is provided by the AGN~\citep{mittal}.  Simulations show that sloshing can provide a non-negligible amount of heat to the cluster core, making it a viable additional heating mechanism~\citep{zuhone}.  The simulations done by \citet{keshet} show that properties of cool core clusters can be explained by the presence of a sloshing spiral, and that all such clusters may contain one.         

In addition to potentially resolving the ``cooling flow'' problem, understanding the substructure in the X-ray gas is important for cosmological studies.  Since clusters are the largest gravitationally bound structures in the Universe, they are a good testbed for cosmological parameters as well as the distribution of dark matter.  Determining these parameters requires accurate knowledge of the mass of the cluster.  Currently, the best estimates for cluster masses come from X-ray data~\citep{review}.  For X-ray mass estimates to be reliable, we must understand in detail the processes occurring in the ICM.  These mass estimates are derived under the assumption that clusters are in hydrostatic equilibrium, but mergers and feedback from AGN can cause departures from the ideal.  Mergers can cause the value of $\sigma_8$ to be overestimated by about 20\%~\citep{randall2002}, and AGN feedback also needs to be taken into account when calculating the masses~\citep{battaglia}.  

Abell 2029 is a relaxed galaxy cluster at a redshift of $z=0.0767$.  It is a Bautz-Morgan type I cluster of galaxies~\citep{abell}.  Using previous \textit{Chandra} observations, \citet{a2029} report a spiral excess in A2029 out to $\sim150$~kpc.  Early observations of  A2029 using the \textit{Einstein X-ray Observatory} and \textit{ROSAT} found a mass deposition rate for the cooling flow of \textit{\.{M}} between $\sim$ 350 and $\sim$ 550~$M_{\sun}$~yr$^{-1}$~\citep{sarazin1992, cf, peres}, while~\citet{a2029} found \textit{\.{M}}$=56^{+16}_{-21}$~$M_{\sun}$~yr$^{-1}$ using a short \textit{Chandra} observation.  Optical observations show that A2029 does not have any indication of [OII] in its spectrum, which is usually associated with a cooling flow~\citep{mcnamara}.

A2029 is host to the radio source PKS1508+059.  The source has the morphology of a C-shaped wide-angle tail (WAT) radio source.  It was originally postulated that WAT shapes were created by a host galaxy's motion through the ICM~\citep{owen}; however later simulations found that the low peculiar velocities of the galaxies combined with the assumed densities and velocities of the particles within the tails themselves were not enough to produce the WAT morphology~\citep{kinetic}.  With this scenario no longer a viable mechanism, it was proposed that cluster mergers could provide the ram pressure necessary to cause the observed morphology.  In the cluster merger scenario, the relative velocities between merging clusters can exceed 1000~km~s$^{-1}$, which would provide the necessary relative velocity to cause the bending.  Observations of some systems showed that WATs were strongly correlated with X-ray substructure, which is indicative of a cluster merger~\citep{burns1994}, and simulations also showed that the morphology could be created during cluster mergers~\citep{roettiger}.  Newer models argue that the bending can occur with lower relative velocities that are consistent with galaxies moving through a cluster ($\sim100 - 300$~km~s$^{-1}$) when the jets have a high flow velocity and low density~\citep{jetha2006}.  It is also possible that gas sloshing could create the WAT morphology.  This has been seen in simulations, even in cases where the gas velocities are not very high~\citep{mendygral2012}.

Throughout this paper, we assume a cosmology with $H_0=70$~km~s$^{-1}$~Mpc$^{-1}$, $\Omega_{\Lambda}=0.7$, and $\Omega_M=0.3$.  At the redshift of A2029 (z=0.0767), this gives a scale of $1''=1.453$~kpc and a luminosity distance $D_L=347.5$~Mpc.  Unless otherwise stated, errors are reported at the 90\% confidence level.

\section{Observations}
We used two \textit{Chandra} pointings totaling 97.7~ks:  19.8~ks from 2000~April~12 (OBSID 891) and  77.9~ks from 2004~January~8 (OBSID 4977).  Both observations were taken in faint (F) mode using the ACIS-S configuration.  The cluster center is located on the S3 chip.  

The data were reprocessed using CIAO version 4.4 and CALDB version 4.4.7.  After the data were reprocessed using \texttt{chandra\_repro} to create new events files, they were then filtered for flares, and background data sets were created using the blank sky background fields reprojected to match the observations.  The S1 CCD chip was used for background flare filtering, as the emission from the cluster takes up the majority of the S3 chip.  To remove the flares, the \texttt{lc\_clean()} routine (as detailed in M. Markevitch's \textit{Chandra Cookbook}\footnote{\texttt{http://cxc.harvard.edu/contrib/maxim/acisbg/COOKBOOK}}) was used.  Filtering was done in the 2.5-6~keV range, with a bin size of 1037.12~s.  The bin size was chosen to reduce the Poisson scatter.  There were no large flares in either data set.  After filtering for flares, no time was removed from the 19.8~ks exposure, while the longer exposure was filtered from 77.9~ks to 76.9~ks.

\section{Images}

\begin{figure*}
\begin{center}
\plotone{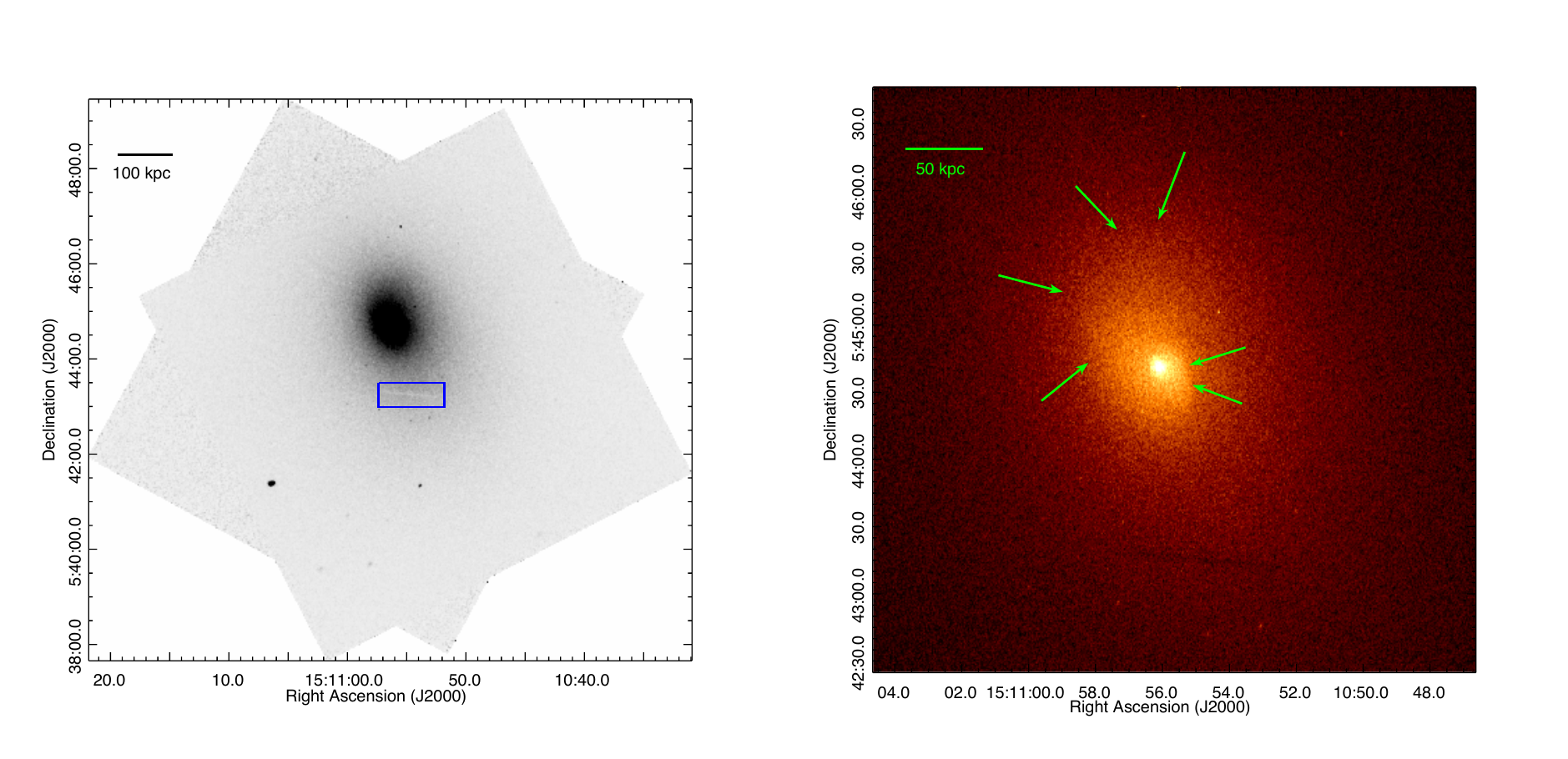}
\caption{The left panel shows a merged, background subtracted, and exposure map corrected \textit{Chandra} 96.7~ks image of A2029 (S3 chip only) in the 0.3--10~keV energy band.  The image has been smoothed with a $1\farcs5$ Gaussian.  Here we emphasize the extended emission of the cluster, which is elongated along the axis from the southwest to the northeast.  The linear negative east-west feature (inside the blue box) to the south of the cluster core at $\alpha=15^h10^m54.6^s$ and $\delta=+05^{\circ}43'13''$  is the absorption shadow of an edge-on foreground spiral galaxy~\citep{spiral}. The right panel shows the central $4\farcm5\times4\farcm5$ of the cluster core in the 0.3--10~keV energy band.  No smoothing has been applied.  The arrows highlight the X-ray surface brightness edges to the east/northeast and the west.}
\label{a2029}
\end{center}
\end{figure*}

A merged, background- and exposure-corrected image of the two exposures using the energy range 0.3--10.0~keV was created, and is shown in the left panel of Figure~\ref{a2029}.  It has been smoothed with a $1\farcs5$ Gaussian.  We restrict our analysis to the S3 chip, which contains the majority of the cluster emission.  The extended cluster emission is visible, as well as several point sources.  The cluster emission shows very little substructure, consistent with A2029 being a relaxed cluster.  It is elongated along the southwest-northeast axis. There is a brightness edge just to the east/northeast of the cluster core, and a second one to the west.  These can be seen in the right panel of Figure~\ref{a2029}, and in the unsharp-masked image described in \S3.1.2.  In all of the images presented (except the spectral maps) one pixel is equal to $0\farcs492$.  

%is v. was here

Background corrections were made using the blank sky background fields.  
For each of the datasets, a reprojected background events file was created, and a background image was created in the 0.3--10~keV range.  The background images were scaled by object exposure time as well as by the ratio of count rates of the S3 chip in the 10--12~keV range after point sources were removed.  The exposure maps for each observation were normalized before being merged.

\subsection{Residual Images}
We used two different techniques---unsharp masking and beta-model subtraction--- to enhance the features in the X-ray image, especially the spiral excess.

\begin{figure}
\begin{center}
\plotone{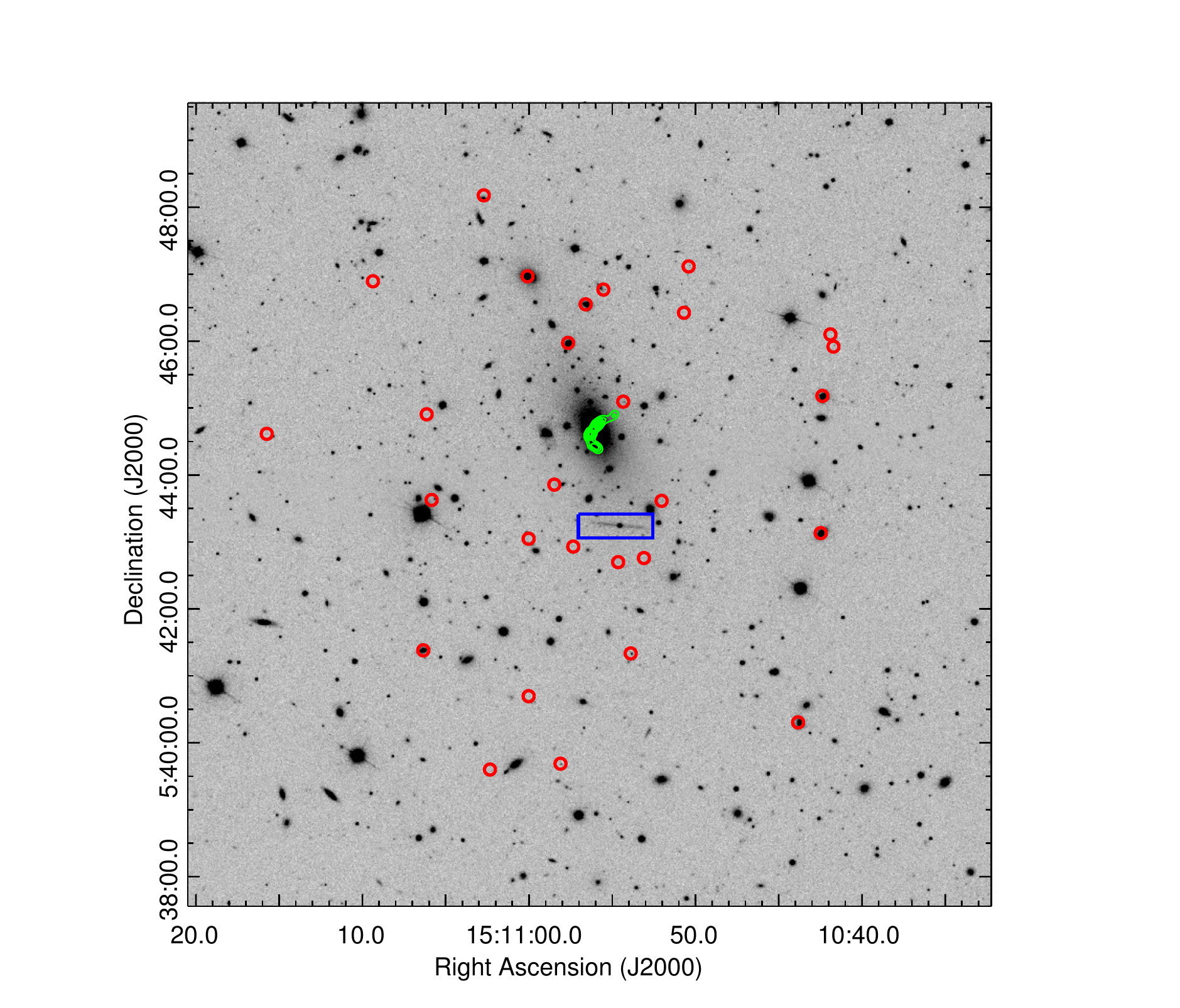}
\caption{SDSS r-band image of A2029.  The field of view matches that of Figure~\ref{a2029}.  The X-ray point sources found by \texttt{wavdetect} are in red, while the radio contours of the central AGN are in green.  The foreground, edge-on spiral galaxy (inside the blue box) that is seen as a shadow in the cluster emission can be seen in the optical image $1\farcm4$ south of the cD galaxy.}
\label{sources}
\end{center} 
\end{figure}

Before making the residual images of the extended, diffuse emission, point sources needed to be detected and removed.  Point sources were detected using the CIAO wavelet detection tool \texttt{wavdetect}.  Wavelet scales of 1, 2, 4, 8, and 16 pixels were used, and the significance threshold was set to $10^{-6}$. The point sources were detected using an unbinned, unsmoothed image (one pixel is equal to $0\farcs492$) for the individual datasets and then merged.  Where there was overlap, the source detection from the $\sim$80 ks image was used.  We visually examined the detected sources and removed those that were likely to be structure in the diffuse gas from the point source list.  There were 34 sources in total.   In Figure~\ref{sources}, an optical r-band image from the Sloan Digital Sky Survey (SDSS) is shown and is matched to the field-of-view of the \textit{Chandra} image in Figure~\ref{a2029}.  The sources are marked with red circles.  

Figure~\ref{sources} also has radio contours displayed to show the central WAT source.  The contours are taken from the VLA Faint Images of the Radio Sky at Twenty-cm (FIRST) survey~\citep{radiocon}.  These contours are used throughout this paper.  Note that the central AGN was not detected as an X-ray point source.  It is likely that the thermal component of the gas from the cool core dominates in this region.  This is discussed further in \S4.1.

\subsubsection{Unsharp Masking}
We created an unsharp-masked image in the 0.3--10.0~keV energy range to enhance faint structure in the diffuse emission.  Following the methods in ~\citet{blanton} and~\citet{fabian2006}, we created three Gaussian smoothed images---one smoothed with a $0\farcs98$ radius kernel with point sources removed, one smoothed with a $9\farcs8$ radius kernel with the point sources removed, and one with a $0\farcs98$ radius kernel with the point sources included.  For the point source free images, the point sources were removed using the tool \texttt{dmfilth}, which replaces pixel values in the region of interest with values interpolated from surrounding regions.  
A point-source free, summed image was created using the two different smoothing scales.  A difference image was then created by subtracting the $9\farcs8$ Gaussian-smoothed source-free image from a $0\farcs98$ Gaussian-smoothed image that still had the point-sources.  The difference image was then divided by the summed image to create the unsharp-masked image.

\begin{figure}
\begin{center}
\plotone{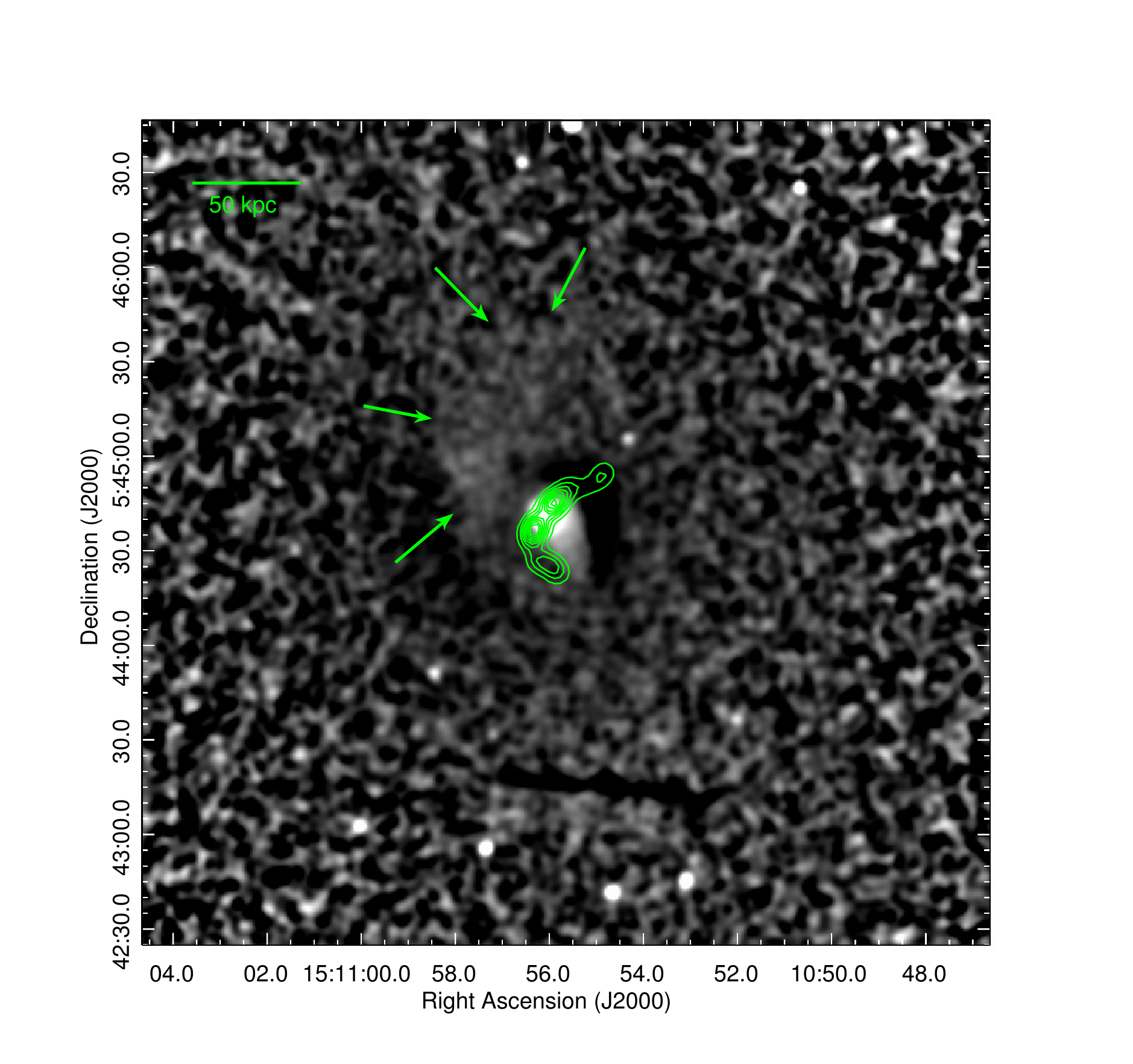}
\caption{The unsharp-masked image of the cluster in the 0.3--10~keV energy band smoothed with a Gaussian kernel of $2\farcs5$ with radio contours overlaid.  There is a prominent surface brightness decrease to the west of the cluster core which the northern lobe of the AGN extends into.  The slight surface brightness excess corresponds to the sloshing spiral, which the green arrows highlight.  The surface brightness excess to the west of the southern radio lobe that extends from the cluster core in between the lobes of the AGN is the very inner edge of the spiral.  The foreground, edge-on spiral galaxy can be seen clearly $1\farcm4$ to the south.}
\label{unsharp}
\end{center}
\end{figure}

Figure~\ref{unsharp} shows the unsharp-masked image with radio contours (taken from the VLA FIRST survey) of the central AGN superimposed.
Unsharp masking reveals the inner region of the spiral as an increase in surface brightness to the north of the cluster center.  There is also a clear decrease in surface brightness to the west of the cluster core, and the northern lobe of the AGN extends into that deficit.  A small surface brightness deficit can also be seen under the southern radio lobe, indicative of a bubble.  This was also seen in~\citet{a2029} at the 4 $\sigma$ level. There is a conspicuous surface brightness excess extending from the cluster core between the lobes of the AGN, which corresponds to the very inner edge of the sloshing spiral.  The foreground, edge-on spiral galaxy noted earlier is seen prominently south of the cluster core.  As discussed in~\citet{blanton}, unsharp masking is better at revealing structure on smaller scales, which is consistent with not being able to see the large-scale spiral structure here clearly; however it does reveal the sharp edges of the inner spiral.
 
\subsubsection{Beta-model Subtraction}
A 2D beta-model was used to fit the surface brightness of the point source-free 0.3-10~keV image.  The model was fitted in \textit{Sherpa} using Cash statistics, and corrections were made for exposure, using the merged exposure map.
%the following is based on the model with the background correction
A constant background of 0.07~counts~pixel$^{-1}$ was also included, as determined from the merged background image.  The best-fitting ellipticity of the emission is 0.25$\pm$0.0012.  The position angle of the semimajor axis of the emission is 21.2$^\circ$$\pm$$0.2^\circ$, measured from north toward east.  These values are consistent with the values found by \citet{a2029}.  The core radius is $27\farcs53\pm0\farcs2$ ($40\pm0.29$~kpc), with a beta index of 0.50$\pm$0.0005.  Errors reported are one-sigma.

\begin{figure}
\begin{center}
\plotone{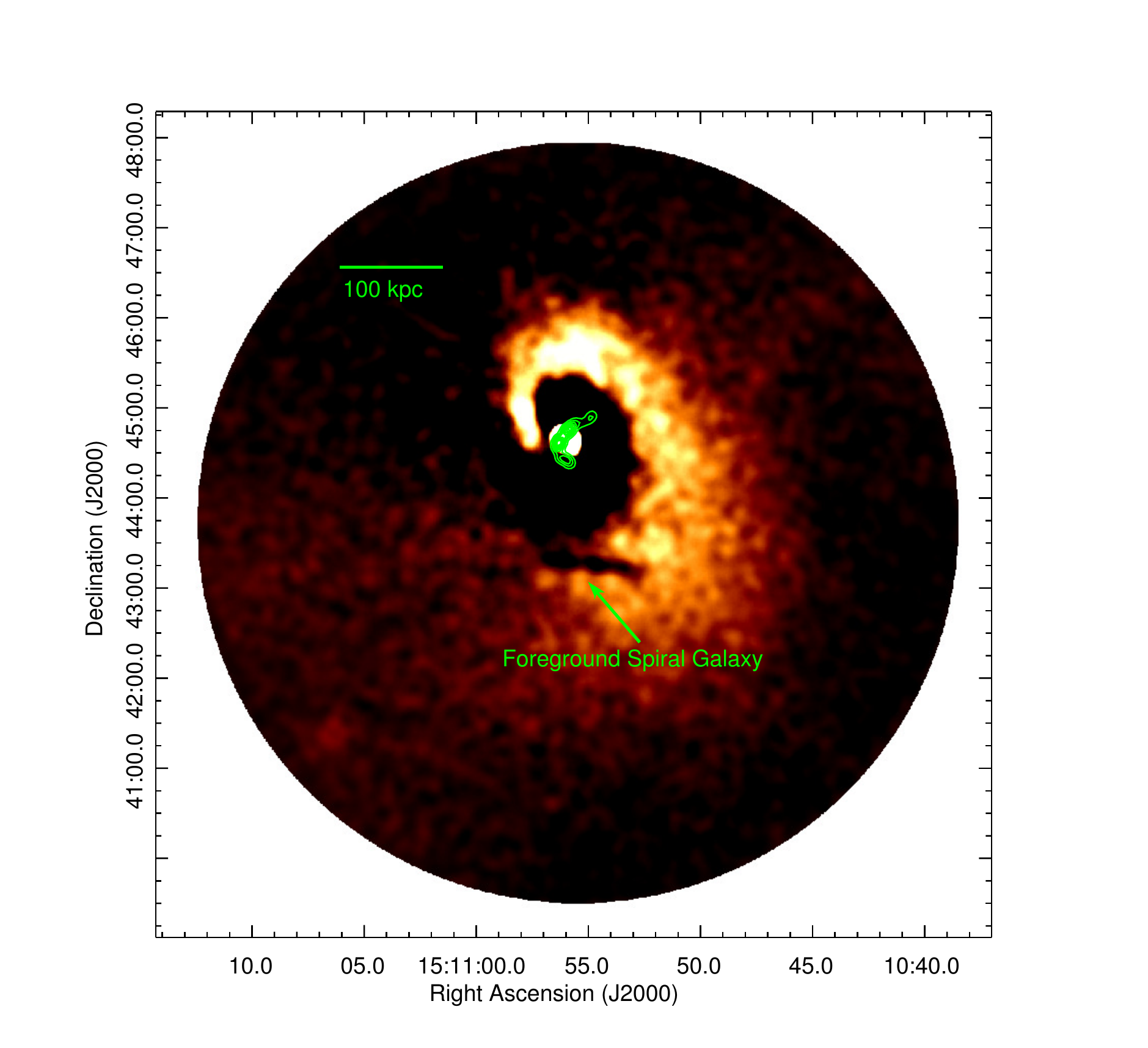}
\caption{Residual image in the 0.3-10~keV energy band after the beta-model subtraction.  The image has been smoothed with a Gaussian kernel of $7\farcs5$.  Contours of the 1.4~GHz radio emission are overlaid.}
\label{beta}
\end{center}
\end{figure}

\begin{figure}
\begin{center}
\plotone{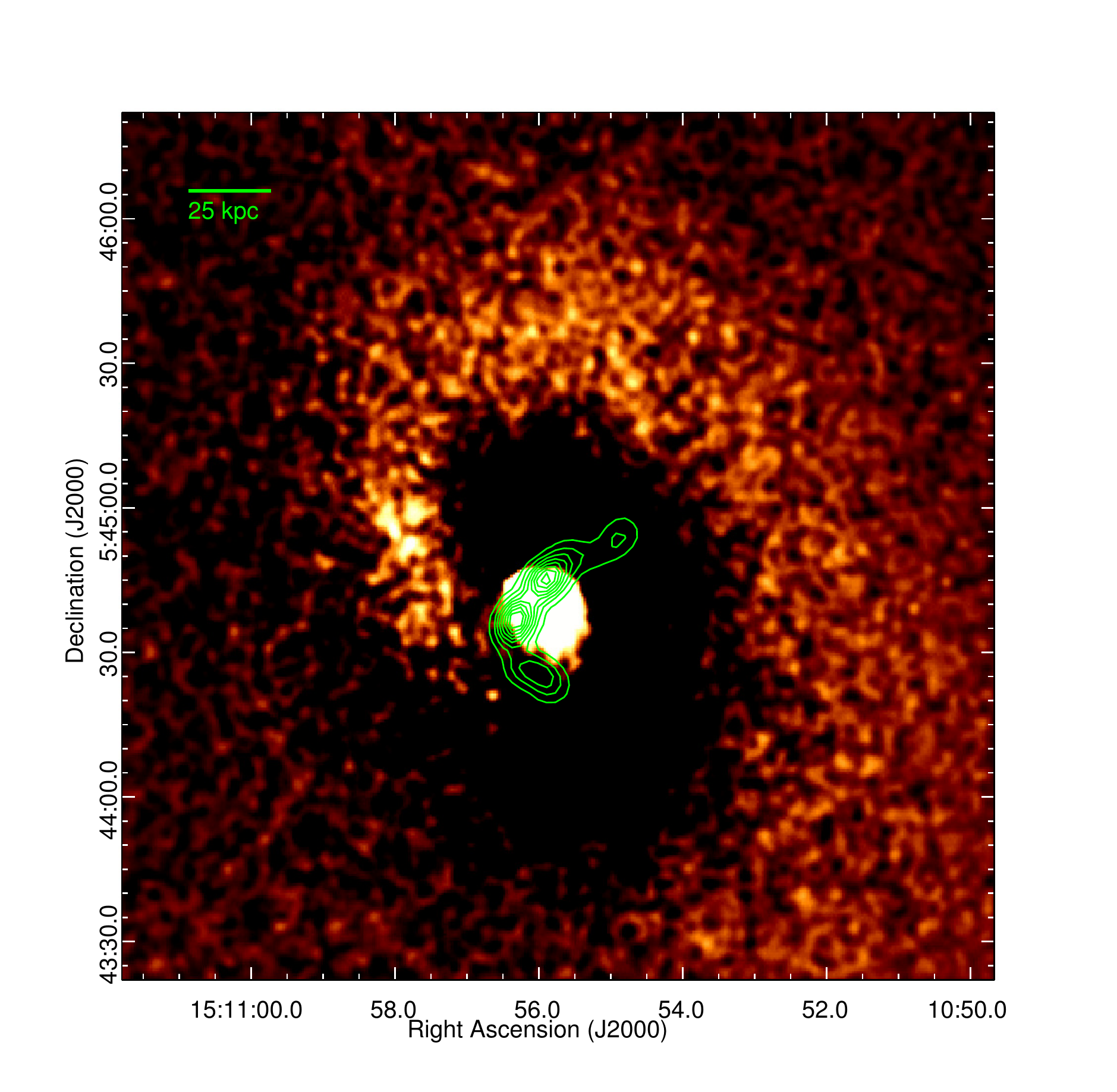}
\caption{Residual image in the 0.3-10~keV energy band of the central $3'\times3'$ region after the beta-model subtraction.  The image has been smoothed with a Gaussian kernel of $2''$.  Contours of the 1.4~GHz radio emission are overlaid.}
\label{beta_zoom}
\end{center}
\end{figure}

The residual image, smoothed with a $7\farcs5$ Gaussian, is shown in Figure~\ref{beta}, with the 1.4~GHz radio contours overlaid.  The spiral excess is clearly visible, and is brightest to the north/northeast of the cluster center.  The spiral extends out to $\sim$300~kpc from the cluster center as measured from the core to the brightest southern edge of the spiral.  Measuring to the very outer southeast edge, the spiral stretches to 400~kpc.  This is the largest \textit{continuous} sloshing spiral observed to date.  Figure~\ref{beta_zoom} shows the central $3'\times3'$ region of the residual image smoothed with a Gaussian kernel of $2"$. Here we can more clearly see the interaction of the sloshing gas with the radio galaxy.   The gas seems to flow along the direction of the spiral, which is predicted by the simulations of \citet{am}, causing an asymmetry in the shape of the radio lobes.  The southern lobe is curved in the direction of the gas flow while the northern lobe is fairly straight.    

\section{Spectral Analysis}
\subsection{Total Spectrum}
The goal of this section is to determine the global properties of the cluster out to a large radius.  We use a radius of 116$''$ (167~kpc), which is the cooling radius found by \citet{sarazin1992} using the \textit{ROSAT} HRI instrument and the radius used for the spectral fits in \citet{a2029}. The analysis was performed with point sources excluded.  The CIAO tool \texttt{specextract} was used to extract the spectra from each observation individually, as well as from the events files for the associated background files.  Weighted response files were created for the data, but not for the background files, since those were to be subtracted.  The spectra were binned with a minimum of 40~counts per bin.  

The spectra from the different observations were fitted simultaneously using the XSPEC software package, version 12.7.0~\citep{xspec}. The fitting was done using the energy range 0.6-7.0~keV.  A systematic error of 2\% was applied and the blank-sky background files were scaled so that the 10-12~keV count rate matched that of the observations.  We explored three different models--a single temperature thermal plasma APEC model, a model with two APEC components, and a model with an APEC component plus a cooling flow model (MKCFLOW).  For each model, we fitted the spectra letting the absorption be both fixed and free.  In the models where the absorption was fixed, it was held at the Galactic value of $3.14\times10^{20}$~cm$^{-2}$~\citep{abs}.  With the cooling flow model, we also explored fixing the lower temperature component to $kT_{low}=0.0808$~keV---the lower limit of the model.   Results of the spectral fitting are shown in Table~\ref{fits}.  The upper and lower error values are the 90\% confidence intervals. 
\begin{deluxetable*}{lcccccc}
%\begin{deluxetable}{lcccccc}
\tabletypesize{\footnotesize}
%\rotate
\tablecaption{XSPEC Fits to the Inner 116{\arcsec} (167 kpc) Radius Region}
\tablewidth{0pt}
\tablehead{
\colhead{Model} & \colhead{$N_H$} & \colhead{$kT_{low}$}
& \colhead{$kT_{high}$} & \colhead{Abundance} &\colhead{$\dot{M}$} & \colhead{$\chi^2/dof$}\\
\colhead{} & \colhead{($\times10^{20}$ cm$^{-2}$)} & \colhead{(keV)} & \colhead{(keV)} & \colhead{($Z_{\sun}$)} & \colhead{($M_{\sun}$ yr$^{-1}$)} &\colhead{}}
\startdata
APEC &(3.14)& \nodata &$7.54^{+0.09}_{-0.09}$ & $0.49^{+0.02}_{-0.02}$ &  \nodata &$802/826=0.97$  \\
APEC & $3.32^{+0.23}_{-0.23}$ & \nodata & $7.48^{+0.12}_{-0.12}$ & $0.49^{+0.02}_{-0.02}$ &  \nodata & $800/825=0.97$ \\
APEC+APEC &(3.14)& $5.97^{+0.56}_{-0.68}$ & $11.16^{+33.7}_{-1.63}$ & $0.52^{+0.03}_{-0.02}$ & \nodata  & $785/824=0.95$ \\
APEC+APEC &$3.33^{+0.23}_{-0.23}$ & $5.98^{+0.69}_{-0.72}$ & $11.31^{+26.4}_{-1.84}$ & $0.51^{+0.03}_{-0.02}$ & \nodata & $783/823=0.95$ \\
APEC+MKCFLOW &(3.14) & $4.28^{+0.74}_{-0.74}$ & $9.68^{+1.45}_{-0.70}$ & $0.53^{+0.03}_{-0.03}$ & $710^{+274}_{-270}$ & $795/824=0.96$ \\
APEC+MKCFLOW &$3.53^{+0.24}_{-0.24}$ & $4.24^{+0.80}_{-0.67}$ & $10.08^{+2.30}_{-1.12}$ & $0.52^{+0.03}_{-0.03}$ & $783^{+371}_{-268}$ & $786/823=0.95$ \\
APEC+MKCFLOW &(3.14) & (0.0808) & $7.54^{+0.09}_{-0.09}$ & $0.49^{+0.02}_{-0.02}$ & $1.43\times10^{-14}$ & $802/825=0.97$ \\
APEC+MKCFLOW &$3.33^{+0.23}_{-0.23}$ & (0.0808) & $7.48^{+0.12}_{-0.12}$ & $0.49^{+0.02}_{-0.02}$ & $1.42\times10^{-14}$ & $800/824=0.97$ \\
\enddata
\tablecomments{Values in parentheses were held fixed during the model fitting.}
\label{fits}
\end{deluxetable*}

We initially fitted the central region using a single temperature APEC model.  This yielded a temperature of $7.54\pm0.09$~keV, which is just outside the range obtained by~\citet{a2029} using a single temperature MEKAL model ($7.27^{+0.16}_{-0.14}$~keV).  Both the APEC model and the MEKAL model are collisional plasma models.  The main difference between the two models is the way that they handle the Fe L line.  The APEC model also adds additional spectral lines to calculate the emission from an optically thin, thermal plasma.  Whereas the MEKAL model uses bundles of closely-spaced emission lines, the APEC model calculates each line individually.  The APEC model also includes more recently updated radiative transition rates and electron collisional excitation rate coefficients.  The single APEC model yielded an abundance of $0.49\pm0.02$~$Z_{\sun}$.  Letting the absorption vary did not change the results of the spectral fit significantly, and yielded a high F-test probability.  This indicates that letting the absorption be a free parameter does not improve the fit.  For both fixed and free absorption, $\chi^2/dof=0.97$.  Leaving the absorption as a free parameter yielded $N_H=3.32\pm0.23\times10^{20}$~cm$^{-2}$, which is consistent with the Galactic value from \citet{abs}. 

Adding a second APEC component to the model improved the spectral fits significantly, yielding an F-test probability of less than 0.02\% compared to the single APEC model.  In this model, the abundances, both temperature components, and both normalization components were tied together for the two observations.  The abundances for both APEC models were also tied together.  With this model, the $\chi^2/dof$ fell to 0.95 for both fixed and free absorption.  Leaving the absorption fixed to the Galactic value, the best-fitting high and low temperatures are $kT_{low}=5.97^{+0.56}_{-0.67}$~keV and $kT_{high}=11.16^{+33.7}_{-1.63}$~keV.  The APEC normalization parameters,
\begin{equation}
K=\frac{10^{-14}}{4\pi D^2_a\left(1+z\right)^2}\int n_H n_e dV,
\end{equation} 
for both components are similar ($K_{low}=2.7^{+1.5}_{-0.9}\times10^{-2} $~cm$^{-5}$ and $K_{high}=2.1^{+0.9}_{-1.3}\times10^{-2}$~cm$^{-5}$).  Since $K$ is proportional to the square of the density, this indicates that one temperature component in not clearly dominant over the other.  The abundance was $0.51^{+0.03}_{-0.02}$~$Z_{\sun}$.  Letting the absorption vary gives $N_H=3.33\pm0.23\times10^{20}$~cm$^{-2}$, $kT_{low}=5.98^{+0.69}_{-0.72}$~keV, $kT_{high}=11.31^{+26.4}_{-1.84}$~keV, and $Z=0.51^{+0.03}_{-0.02}$~$Z_{\sun}$.   In~\citet{a2029}, the two component MEKAL model yielded a high temperature of $kT_{high}\sim7.5$~keV and a low temperature of $kT_{low}\sim0.1$~keV within the same spatial region.  Our high temperature value is closer to the temperature found by~\citet{V05}, who also used \textit{Chandra}.  They find that between $\sim100-300$~kpc the cluster temperature rises to $\sim9$~keV.  The region we explore extends out to approximately 170~kpc, so the area of increased temperature is included.

Because there is evidence of a cooling flow in A2029, we also fitted a model including a cooling flow component (MKCFLOW) with a single-temperature APEC model to account for gas outside the cooling region.  The MKCFLOW model is based on one presented by \citet{mkcflow}, who found that two isothermal plasmas did not provide a good physical model for cooling in clusters.  This model assumes that the mass flow rate is consistent throughout the cooling flow.  In this scenario, the high temperature ($kT_{high}$) component of the cooling flow model was tied to the temperature of the APEC model.  As with the APEC+APEC model, the abundances of each dataset and each model were tied together.  Keeping the absorption fixed, the MCKFLOW+APEC model gives best-fit temperatures $kT_{high} = 9.68^{+1.45}_{-0.70}$~keV and $kT_{low}=4.28^{+0.74}_{-0.74}$~keV.  The best-fit abundance is $0.53\pm0.03$~$Z_{\sun}$.  The cooling flow mass deposition rate is \textit{\.{M}}$=710^{+274}_{-270}$~$M_{\sun}$~yr$^{-1}$.  Letting the absorption vary gives a column density of $N_H=3.53\pm0.24\times10^{20}$~cm$^{-2}$, $kT_{high}=10.08^{+2.30}_{-1.12}$~keV, $kT_{low}=4.24^{+0.80}_{-0.67}$~keV, and an abundance of $0.52\pm0.03$~$Z_{\sun}$.  With the absorption free, \textit{\.{M}}$=783^{+371}_{-268}$~$M_{\sun}$~yr$^{-1}$. Comparing the MKCFLOW+APEC model with absorption free to a single-temperature APEC model (with absorption free and fixed) gives an F-test probability of $\sim0.1\%$, which indicates that it is a significant improvement.  Comparing the two MKCFLOW+APEC models (one with the absorption free and the other with the absorption fixed), an F-test shows that the one with the free absorption model is a significant improvement, yielding a probability value of $\sim0.7\%$.

The high temperatures in the cooling flow model are consistent with the values from~\citet{V05}, and the low temperature is consistent with the results from other clusters that show gas cools to about one-half to one-third of the ambient temperature~\citep{peterson}.  We also fitted an APEC plus cooling flow model where $kT_{low}$ was fixed to $kT_{low}=0.0808$~keV, which is the hard lower limit in XSPEC, to compare with~\citet{a2029}.  In this model, the best-fit APEC values matched those of the corresponding (i.e. absorption fixed or free) single-temperature APEC model discussed above and seen in Table~\ref{fits}.  With this fixed lower limit, the cooling flow rate was $\sim$ $10^{-14}$~$M_{\sun}$~yr$^{-1}$, making the MKCFLOW component negligible.  From this we can infer that a significant fraction of the gas in this region is not cooling to very low temperatures.  

The best-fit cooling flow rate of \textit{\.{M}}$=783^{+371}_{-268}$~$M_{\sun}$~y$r^{-1}$ determined here is higher than previously found values of between 350 and 550~$M_{\sun}$~yr$^{-1}$~\citep{sarazin1992, cf, peres}; however, there is overlap in the confidence intervals of our value and the higher previous value of $556^{+215}_{-93}$~$M_{\sun}$~yr$^{-1}$, found using \textit{ROSAT}~\citep{peres}.  These values are significantly higher than the rate of 56~$M_{\sun}$~yr$^{-1}$ found by~\citet{a2029} using the same region for spectral extraction. \citet{a2029} also find a $kT_{high}/kT_{low}>50$, whereas we find $kT_{high}/kT_{low}\sim2.4$, which is more in agreement with values seen in other clusters~\citep{peterson}.  We explore this discrepancy further in \S7.

\subsection{Central Source}

We would like to determine whether or not the central AGN can be detected spectrally.  To do this, we extracted the spectrum of the central source using an aperture of $2\farcs0$ (2.9 kpc) centered on the coordinates of the radio galaxy core as reported by \citet{taylor} ($\alpha=15^h10^m56.82^s$, $\delta=+05^{\circ}44'41.26''$, J2000).  The background was determined locally from an annulus centered on the source with an inner radius of $2\farcs0$~(2.9~kpc) and an outer radius of $5\farcs0$~(7.3~kpc).  There were approximately 1100 net counts in the region.

We fitted both a non-thermal power-law and a single-temperature APEC model with absorption both fixed and free to the region.  Parameters from each dataset were tied together.  The best-fitting photon index for the power-law model is $\Gamma=2.01\pm0.09$, which is slightly higher than the photon index of $\Gamma\sim1.7-1.8$ seen in 90\% of radio-loud AGN~\citep{gamma}.  Letting the absorption vary improves the power-law model (lowering the $\chi^2/dof$); however, the value of $\Gamma$ rises to $\Gamma=3.10^{+0.35}_{-0.30}$.  This is well above the average range of values seen in other radio-loud galaxies.  The single-temperature APEC model is a better fit to the region.  The best-fit model for the central region is the single-temperature APEC model with the absorption free to vary; however it is not a significant improvement over fixed absorption as the F-test probability is $\sim39\%$.  The APEC model with free absorption yields an absorption of $N_H=6.38^{+6.23}_{-5.49}\times10^{20}$~cm$^{-2}$, a temperature of $kT=2.47^{+0.44}_{-0.36}$~keV, and an abundance of $Z=1.29^{+0.92}_{-0.55}$~$Z_{\sun}$.  Neither an APEC+APEC model nor an APEC+PO model was an improvement over the best-fitting single-temperature APEC model.  Based on our best-fit model there is no detection of non-thermal emission from an AGN.  This is consistent with \citet{a2029}, who found that the best fit model was a single-temperature MEKAL model.  They were unable to find good fits with a power-law component with a photon index $\Gamma<3.4$.

\subsection{X-Ray Spectral Maps}
Spectral maps were created to examine the distributions of temperature, entropy, and abundance in the inner 250"~(370~kpc) region of the cluster, following the technique outlined in~\citet{randall2008,randall2009}.  Spectra were extracted from the 0.6-7 keV energy range.  For the temperature and entropy maps, we used a minimum of 3,000~net counts per fit, and for the abundance map we used a minimum of 10,000~net counts per fit to improve the errors.  The maps are binned such that each pixel is 10''.  The radius of the extraction region for each pixel is allowed to grow until the minimum number of net counts is contained inside the extraction region.  In the central region where the surface brightness is high, there is little-to-no overlap between the extraction regions.  There, each 10" pixel represents a spectral fit, while at the edge of the field the surface brightness is lower and the neighboring regions overlap significantly.  As many as 15~pixels are contained in an extraction region at the edge of the field.  Point sources were excluded from the spectral fits, as was the region that includes the foreground, edge-on spiral galaxy.  The column density was fixed to the Galactic value of $3.14\times10^{20}$~cm$^{-2}$, but elemental abundance values were allowed to vary.

\begin{figure}
\begin{center}
\plotone{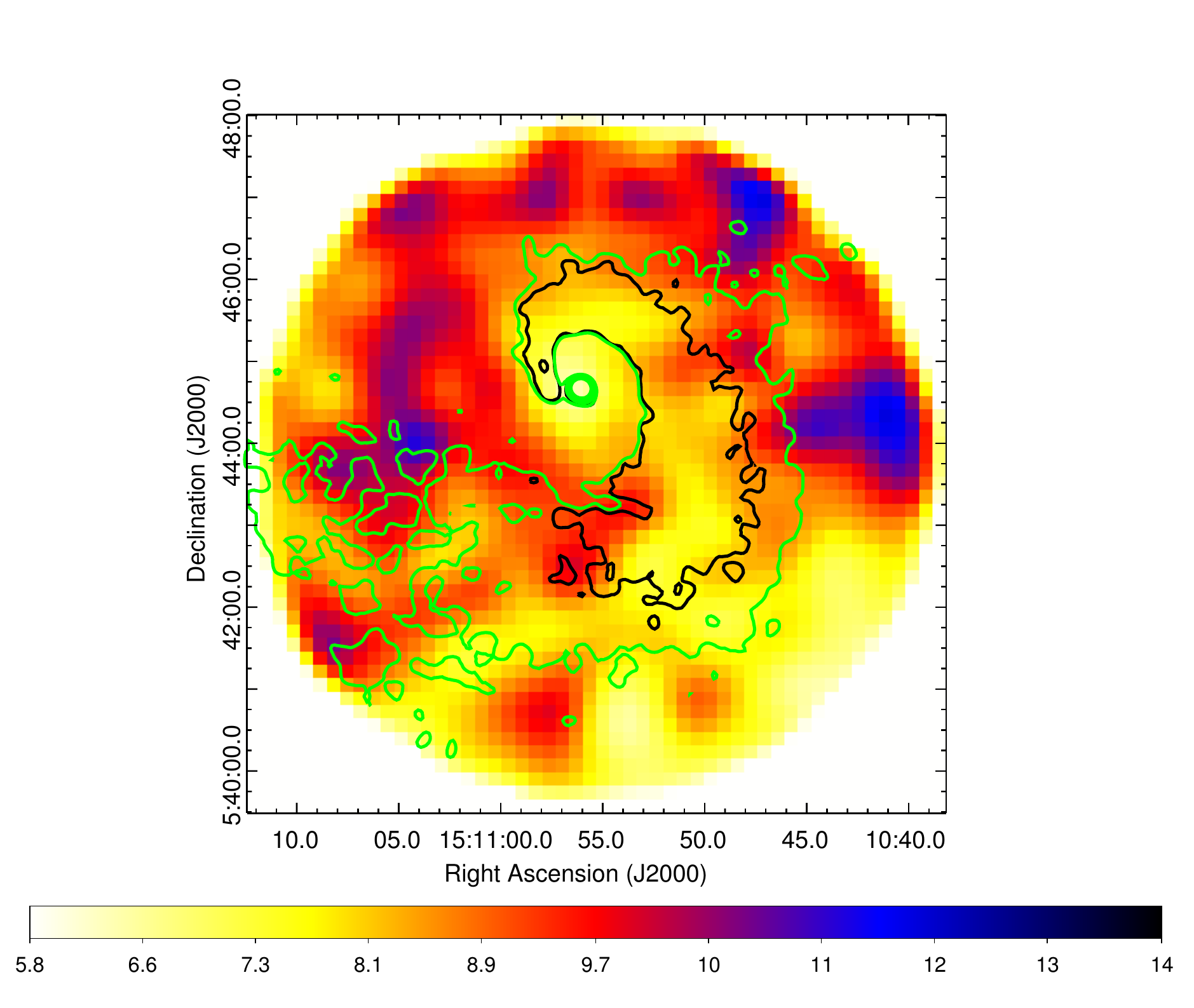}
\caption{Temperature map of A2029 with contours of the brightest part of the spiral overlaid in black and contours of the full extent of the spiral overlaid in green.  The spiral is coincident with the cooler parts of the cluster.  Each temperature was calculated using a minimum of 3000~net counts in the energy range 0.6-7~keV.  The regions were allowed to grow from their minimum binning of 10$''$ so that each region would meet the minimum count requirement.  The image has been smoothed with a Gaussian with a kernel radius of $30''$.  The scale bar is in \textit{kT} with units of keV.}
\label{tmap}
\end{center}
\end{figure}

Figure~\ref{tmap} shows a temperature map of A2029 with contours of the brightest part of the spiral overlaid in black and the full extent of the spiral overlaid in green.  The spiral feature is coincident with the cooler parts of the cluster, as predicted in the sloshing model.  Errors are, on average, between 10\%-15\% across the map. They range from less than 10\% in the center of the cluster where the count rate is highest, to approximately 25\% at the outer edge of the map.

Along with cooler temperatures, we expect the region with the spiral excess to show higher abundances, since the sloshing displaces the metal-rich gas from the cluster core and brings it into contact with the more metal-poor ICM outside the cluster center.
\begin{figure}
\begin{center}
\plotone{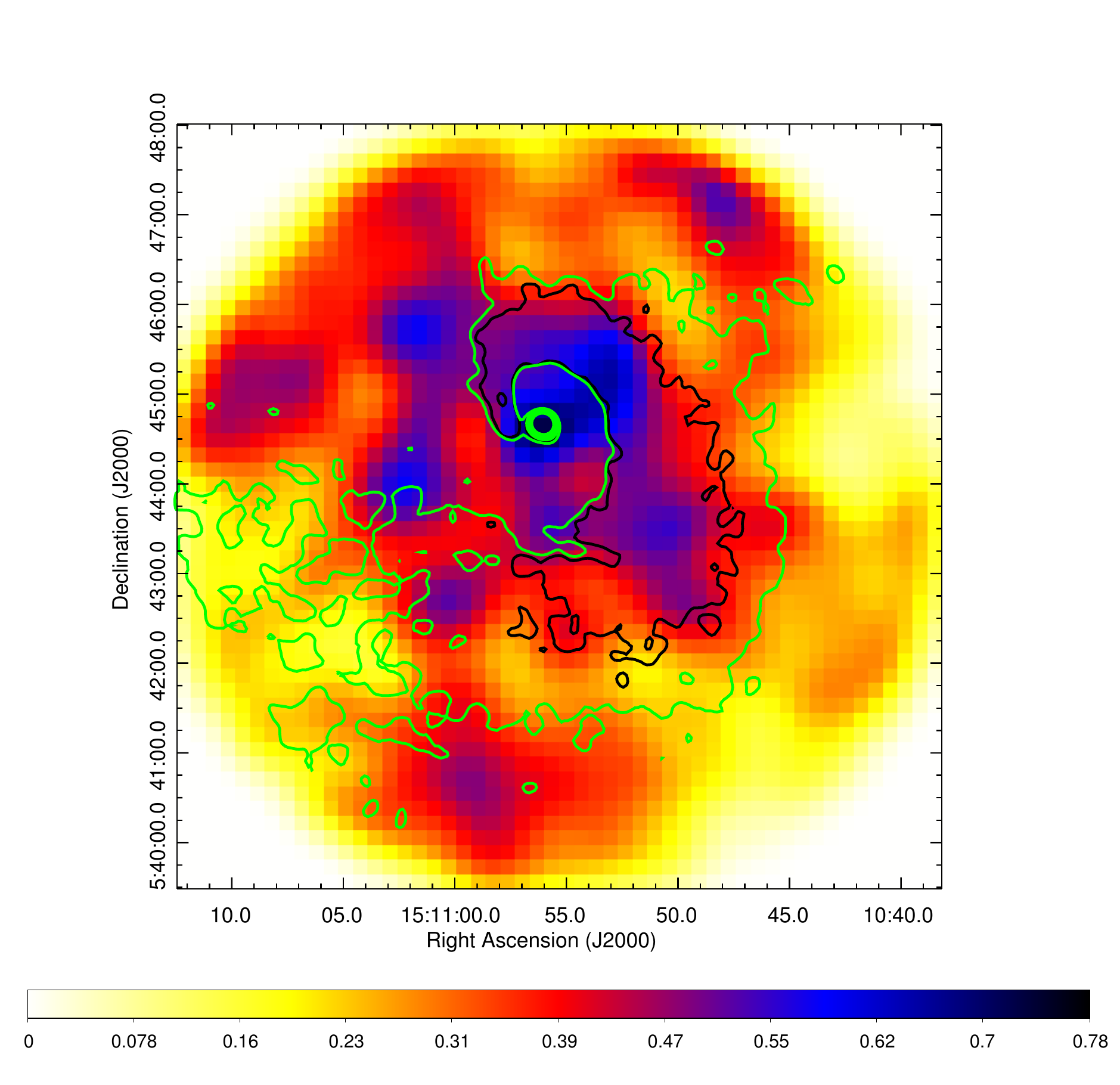}
\caption{Abundance map of A2029 with contours of the brightest part of the spiral overlaid in black and contours of the full extent of the spiral overlaid in green.  The brightest part of the spiral appears coincident with a region of high abundance, although high abundance regions are are also seen elsewhere in the cluster.  Spectral fits were performed using a minimum of 10,000~net counts per region.  This map has been smoothed with a Gaussian with a kernel radius of $30''$.  The scale bar shows abundance in units relative to the solar abundance.}
\label{abun}
\end{center}
\end{figure}
Figure~\ref{abun} shows that the contours tend to trace the areas of highest abundance, most clearly in the bright, northern region.  This is consistent with the idea that the higher metallicity gas in the center of the cluster is being displaced, creating the sloshing spiral.  While the errors are large, the abundance map can be used as a visualization tool.  The errors start at 20\% at the lowest in the center of the map and reach up to 50\% towards the edge of the field.  Within the spiral region, the errors range from 20\% to 40\%.

We also construct a pseudo-entropy map.  Pseudo-entropy is defined as $kT(K/A)^{-1/3}$, where $K$ is the APEC normalization, and $A$ is the on-chip area of the extraction region.  Since $K\propto n_e^2$, the pseudo-entropy is proportional to $k T n_e^{-2/3}$.  The combination of low temperature and high density in the spiral enhances the entropy and there is a noticeably lower entropy in the spiral region.
\begin{figure}
\begin{center}
\plotone{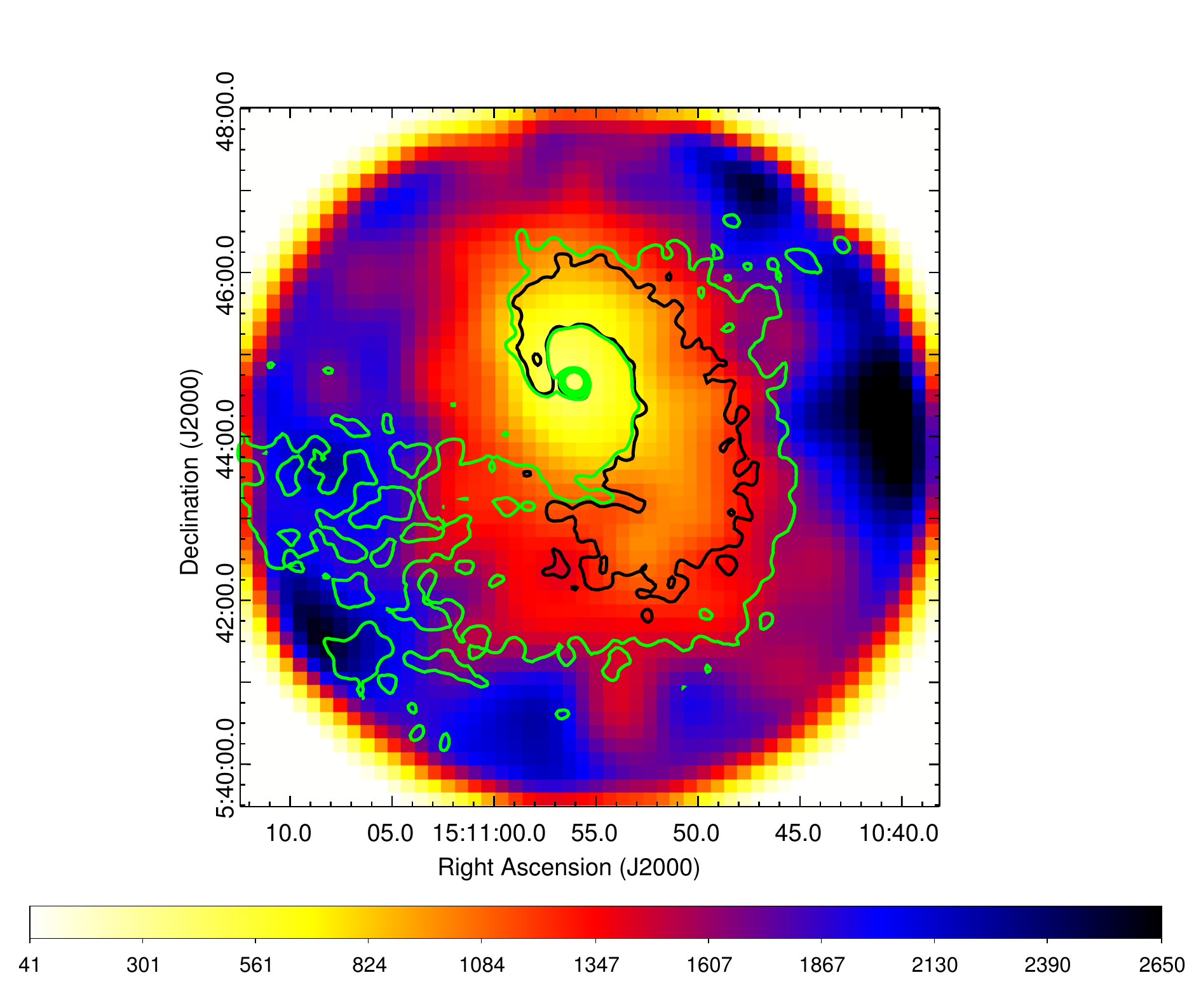}
\caption{Psuedo-entropy map of A2029 with contours of the brightest part of the spiral overlaid in black and contours of the full extent of the spiral overlaid in green.   The scale bar is in arbitrary units.  The part of the cluster with the lowest entropy is coincident with the spiral feature, and the entropy increases outward. The image has been smoothed with a Gaussian with a kernel radius of $30''$.}
\label{entropy}
\end{center}
\end{figure}
This is seen in Figure~\ref{entropy}, where the lowest entropy portion of the cluster follows the spiral structure.  There is also an overall trend of entropy increasing away from the center of the cluster.

The spectral maps show that the distribution of temperature, abundance, and entropy are correlated with the spiral structure.  We explore this structure in more detail in the following section.
  
\section{The Sloshing Spiral Feature}

The spiral feature visible in Figure~\ref{beta} is similar in morphology to the spiral features induced by sloshing seen in simulations of cluster-cluster mergers originally described in~\citet{am}, and more recently in e.g.~\citet{zuhone} and~\citet{zuhone2012}.  The simulations in \citet{zuhone2012} included magnetic fields to stabilize the cold fronts against Kelvin-Helmoltz instabilities (KHIs), which give the appearance of ragged edges, in contrast with the smooth edges that are observed in many systems. Some systems, however, such as A496~\citep{a496} and the merging groups NGC7618 and UGC12491~\citep{kh} do show distortions that are consistent with KHIs.  The prevalence of smooth edges may be due to viewing the cold fronts in projection~\citep{roediger}.  

In the simulated mergers, the cool, central gas in the cluster is displaced to larger radii, which is known as sloshing.  The sloshing sets up cold fronts, and the spiral structure results when the merger is off-center, transferring angular momentum.  We would expect the brightness excess seen in Figure~\ref{beta} to also be seen as excesses in the surface brightness profile of the region including the spiral.  The temperature profile of these regions should show cooler gas coincident with the regions of surface brightness excess.  A list of global sloshing features (e.g. brightness edges, large-scale asymmetry, temperature and metallicity structure) can be found in \citet{a496}.  

\begin{figure}
\begin{center}
\plotone{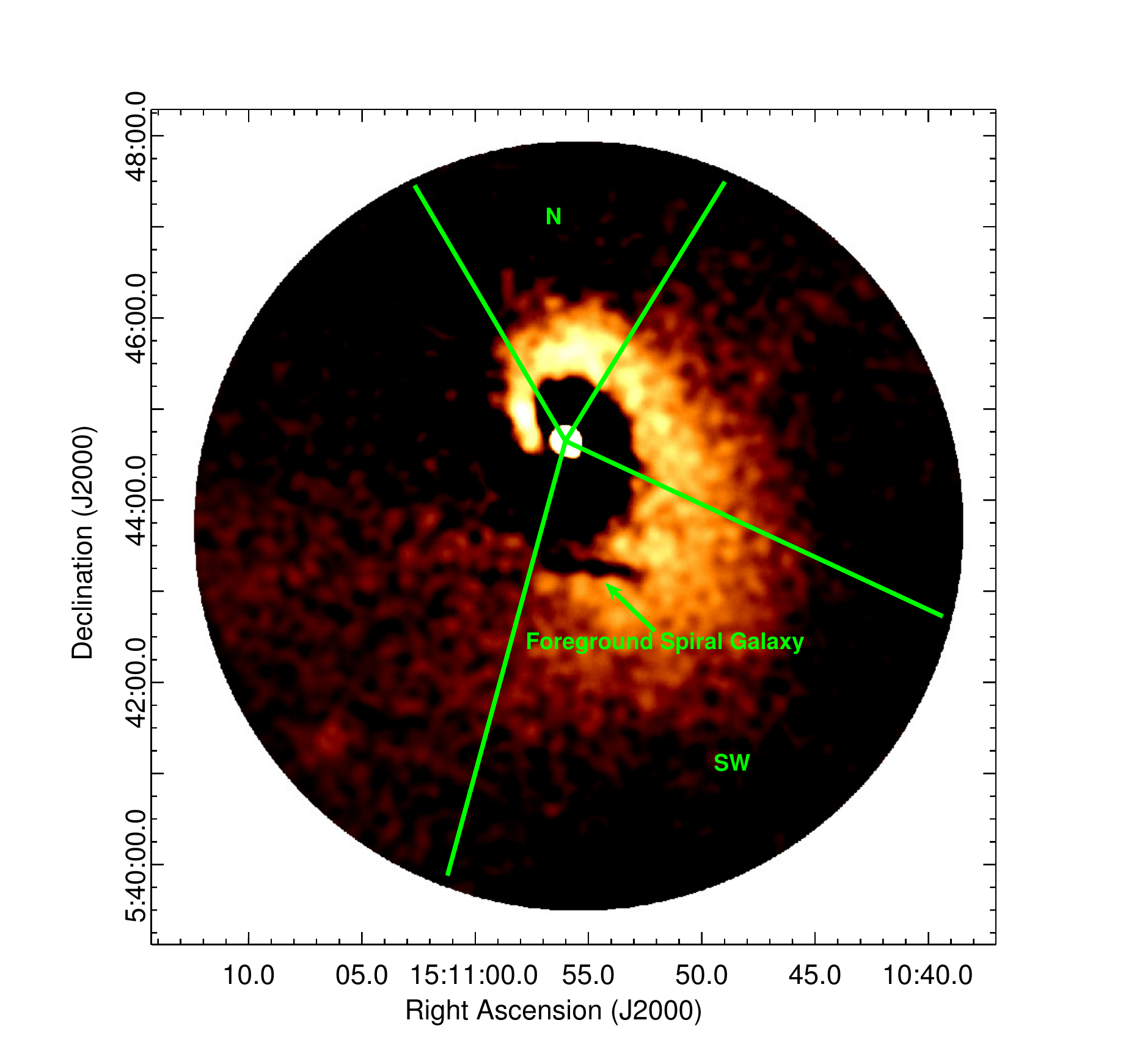}
\caption{The two regions used for creating surface brightness, temperature, density, and pressure profiles overlaid on the residual image of the spiral feature.}
\label{wedge}
\end{center}
\end{figure}

\begin{figure*}
\begin{center}
\plotone{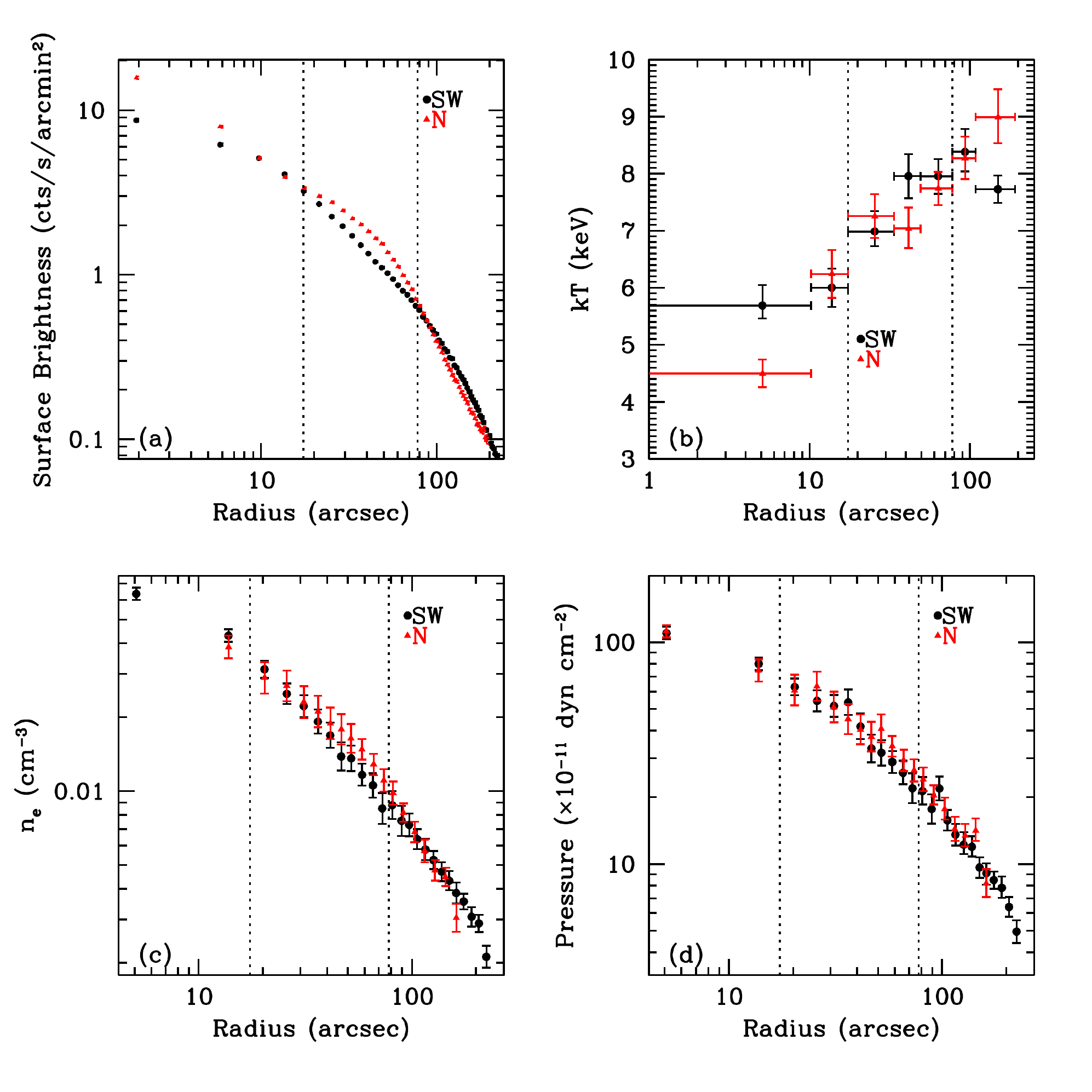}
\caption{Surface brightness profile (a), projected temperature profile (b), density profile (c), and pressure profile (d) using the sectors shown in Figure~\ref{wedge}.  In the N region, the surface brightness excess stretches from $\sim20''$~($\sim29$~kpc) to $\sim80''$~($\sim116$~kpc), while in the SW region the surface brightness excess extends past $\sim80''$~($\sim116$~kpc). The inner dashed line marks the beginning of the spiral excess to the north.  The outer dashed line marks the other edge of the spiral in the north and the inner edge of the spiral in the southwest.  Error bars are one-sigma.}
\label{panel}
\end{center}
\end{figure*}

We have extracted surface brightness profiles from wedges in two directions, north (N) and southwest (SW), as shown in Figure~\ref{wedge}.
The N region includes the inner part of the spiral, while the SW region includes the outer part of the spiral.  Due to the placement of the cluster on the chip, we were able to go out to a larger radius when exploring the properties of the SW region.  The surface brightness profile obtained from the original merged image is shown in Figure~\ref{panel}a.  Many of the error bars are smaller than the symbols used.  The excess in the N region is seen between $\sim20''$~($\sim29$~kpc) and $\sim80''$~($\sim116$~kpc), while the excess emission in the SW region extends beyond $\sim80''$~($\sim116$~kpc).  We examined a 2830 arcsec$^2$ circular region in the northern part of the spiral where there was a surface brightness excess seen in the surface brightness profile and compared it with an identical region at the same distance from the cluster center directly south of the core where no surface brightness excess was observed.  We did the same for examining the SW excess (as compared to the north) at larger radii using a 7850 arcsec$^2$ circular region.  We find a 24\% excess of counts in the brightest part of the spiral to the north of the cluster center compared to the same radial range in the SW, corresponding to 44$\sigma$.  In the southwest region, corresponding to the outer part of the spiral, there is a 22\% excess of counts compared to the same radial range in the north, corresponding to 26$\sigma$.

\begin{figure*}
\begin{center}
\plotone{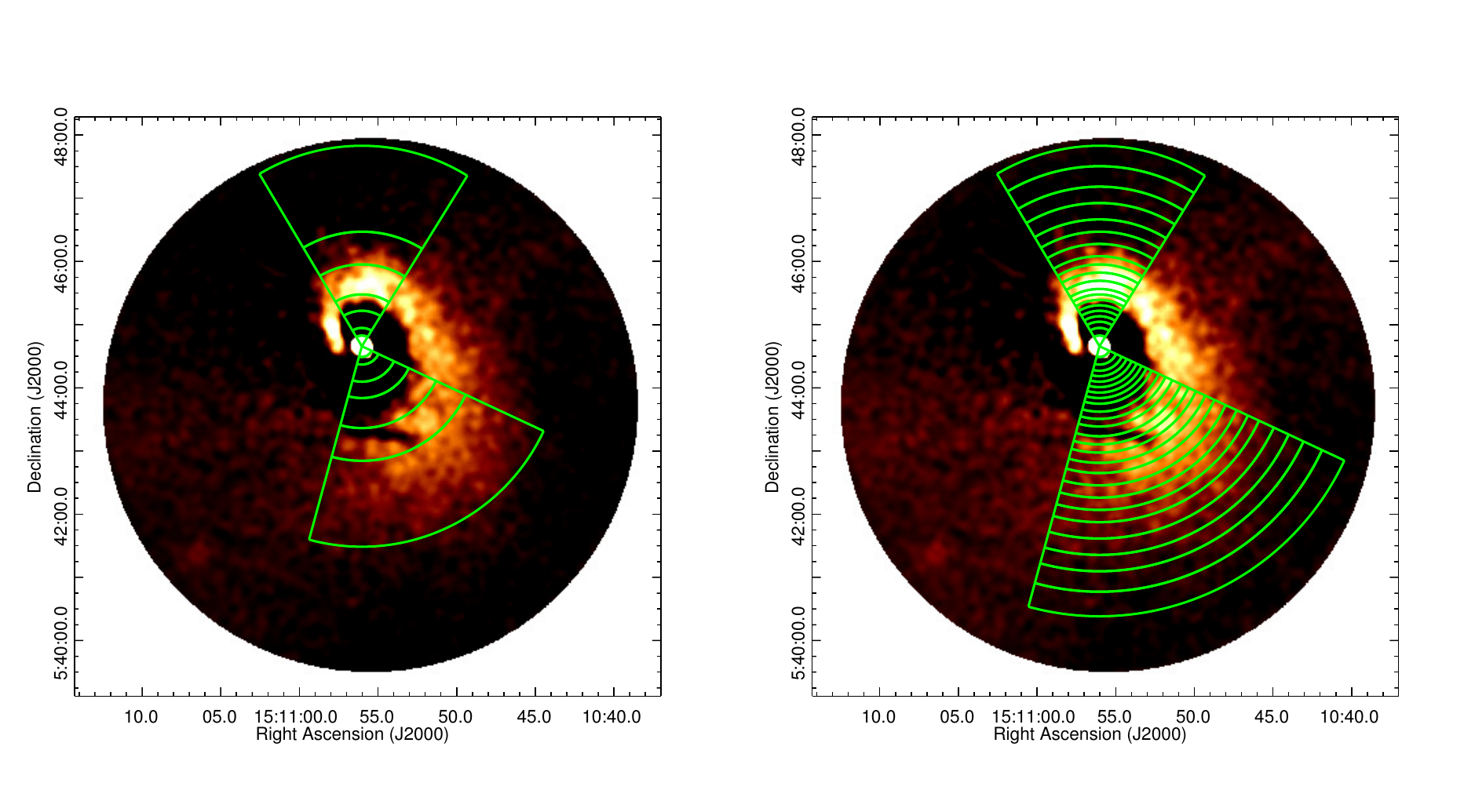}
\caption{Annuli used for the temperature profile (shown in the left panel) and for the density and pressure profiles (right panel) shown in Figure~\ref{panel}.}
\label{annuli}
\end{center}
\end{figure*}

In addition to determining the surface brightness profile in these two directions, we also measure the spectral properties.  For the temperature profile, we have extracted spectra from the same two wedges, but using fewer annuli, using the method described in \S4.  These regions are shown in the left panel of Figure~\ref{annuli}.
The spectra were fitted using an APEC thermal plasma model with the absorption fixed to the Galactic value.  Elemental abundances were allowed to vary.   
The temperature profile is shown in Figure~\ref{panel}b.  The regions used to extract the spectra are larger than the regions used in the surface brightness profile to minimize the error bars.  Each region has a minimum of 10,000~net counts.  In the region to the north that corresponds to the surface brightness excess, the temperature is cooler than the SW region, where there is no surface brightness excess at those radii.  This temperature relationship is inverted for the outer portion of the spiral.  There the temperature in the SW region (where there is a surface brightness excess) is cooler than points at the same radii in the northern region (where the surface brightness is lower).  The surface brightness excess to the north is the greatest at $r\sim40''-50''$~($r\sim58-73$~kpc), which is where we see the gas is most significantly cooler.  For the southwest, the surface brightness is greatest at $r\geq80''$~($r\geq116$~kpc), and the temperature is correspondingly cooler here than to the north.  The relationship between the two regions is as expected in a sloshing model, where cool gas in the cluster core has been displaced to larger radii in a spiral distribution.

Figure~\ref{panel}c shows the electron number density profile of the two regions, with the annuli shown in the right panel of Figure~\ref{annuli}.  The density was calculated by deprojecting the X-ray surface brightness, assuming that the emissivity is constant within spherical shells, following the method described in \citet{blanton09}, \citet{wong}, as well as \citet{kriss}.  The surface brightness is proportional to $n^2T^{1/2}$.  Because of the weak temperature dependence, the density is not a very sensitive function of the temperature.  The projected spectral fits are used to convert count rates to flux, which can be combined with the emissivity to calculate density.  Combining the density with the temperature (which was found by fitting each region with a single APEC model in XSPEC), we determine the pressure.  Each region contains at least 10,000~counts.  In the region of the inner spiral, the density is higher to the north than the same radial values in the SW region, where there is no surface brightness excess.  The density becomes noticeably enhanced in the north at about 40$''$~($\sim29$~kpc), which is where the surface brightness excess is largest (as seen in Figure~\ref{panel}a).  It is more difficult to see this effect in the outer region of the spiral.  The difference in surface brightness is not as large, and there are also fewer points for comparison.  

Figure~\ref{panel}d shows the pressure profile of the two regions.  The same regions used to calculate the density across the spiral were used to calculate the pressure.  The density combined with the projected temperature of each radius was used to determine pressure.  The pressure is similar at corresponding radii in both the N and SW regions of the spiral.  We see no significant jumps in the pressure profile that would correspond to shocks.  The pressure profile is smooth across the edges of the spiral feature, which is consistent with cold fronts.
 
\section{Bending of WAT Lobes Through ICM Sloshing}

The radio galaxy at the center of the cluster is affected by the spiral feature, and vice versa.  The radio galaxy appears to have inflated bubbles with its lobes, as evidenced by the way that the spiral does not smoothly connect to the center of the cluster in the south, and the deficit of X-ray emission at the position of the NW lobe.  It is also likely that the spiral is affecting the shape of the radio lobes, giving them their wide-angle tail morphology.  The lobe that has evacuated the cavity to the south appears to be bent more than the other lobe.  This type of asymmetry was found in simulations by~\citet{mendygral2012}.  They performed three-dimensional magnetohydrodynamical simulations to explore the interaction between AGN and ICM ``weather" such as sloshing and found that ICM flows could cause asymmetries between the AGN lobes and jets even in seemingly relaxed clusters.  The motion of the ICM is most likely at an angle to the jet axis, as the AGN morphology is a cross between the two limiting cases---ICM motion aligned with the jet axis and ICM winds orthogonal to the jet axis---discussed in~\citet{mendygral2012}.  We see that there is a wide-angle tail morphology, but the southern jet is more bent than the northern jet.  This kind of morphology is seen in the simulations when the ICM winds are orthogonal to the jet axis.  The northern jet is more elongated than the southern jet, which is seen in the simulations when the ICM winds are aligned with the jet axis.  The southern lobe is bent towards the southwest.  The gas is expected to flow inwards along the spiral (see the velocity vectors in \citet{am}, specifically their Figure~7), so the displacement of the lobe is consistent with the flow of gas along the portion of the spiral that runs parallel to the southern lobe.  This kind of distortion can help provide information about the velocity structure of clusters where it is not possible to do direct measurements.

Using a method similar to the one described in \citet{douglass, douglass2011}, we estimate the relative velocity between the ICM and radio galaxy. If we assume pressure balance between the radio lobes and the ICM, we can equate the ram pressure of the buoyant radio lobes to the dynamic ram pressure of the ICM as it sloshes past the AGN, which gives
\begin{equation}
\frac{\rho_r v^2_r}{r_c}=\frac{\rho_{ICM}v^2_g}{r_r},
\label{euler}
\end{equation}
where $r_r$ is the radius of the lobe, $v_r$ is the velocity of the plasma in the lobe, $\rho_r$ is the mass density of the lobe, $\rho_{ICM}$ is the density of the ICM (where $\rho_{ICM}=1.92n_e\mu m_p$; $\mu$ is the mean atomic mass of the cluster and is assumed to be 0.6 and $n_e$ is found from the density profile), $v_g$ is the velocity of the galaxy relative to the ICM, and $r_c$ is the radius of curvature of the source.  Using the kinetic model of \citet{kinetic}, the condition used to determine the internal density required to produce the observed luminosity of the lobes is
\begin{equation}
L_{rad,l} = \frac{\epsilon\pi r^2_r \rho_r v^3_r}{2}.
\label{luminosity}
\end{equation}
This model assumes that some fraction ($\epsilon$) of the kinetic energy is converted into the observed radiation.  Combining this with Equation~\ref{euler}, the velocity of the galaxy relative to the ICM is:
\begin{equation}
v_g = \left(\frac{2 L_{rad,l}}{\epsilon\pi\rho_{ICM}v_r r_c r_r}\right)^{1/2}.
\label{velocity}
\end{equation} 
Here, $L_{rad,l}$ is the radio luminosity of the lobe.

To determine the efficiency with which the kinetic energy of the outflow is converted into radiation, we need to determine the mechanical power of the source and the total radio luminosity of the source ($\epsilon=L_{rad}/(E_{kin}/t)$).  The kinetic energy of the AGN goes into evacuating cavities in the ICM and, in the case of relativistic jets, $E_{kin}=4PV$~\citep{Ekin}.  Assuming that the AGN is in pressure equilibrium with the ICM, we can estimate the pressure from our pressure profile and find that at the outer edge of the southern lobe, $P=  6.4\times10^{-10}$~dyn~cm$^{-2}$.  To determine the volume, we estimate a circular radius for the southern lobe of $4\farcs7$~(6.9~kpc).  To determine the power, we assume a typical AGN repetition rate of $5\times10^7$~years~\citep{birzan}.  We multiply this power by two to account for both lobes.  To find the radio luminosity, we use the following equation:
\begin{equation}
L_{rad}=4\pi D_L^2 S_{\nu_0}\int_{\nu_1}^{\nu_2}\left(\frac{\nu}{\nu_0}\right)^{-\alpha}d\nu.
\label{radio}
\end{equation} 
We find the total radio luminosity to be $3.8\times10^{42}$~erg~s$^{-1}$, using a spectral index of $\alpha=1.52$, a reference flux of 511~mJy at 1490~MHz~\citep{taylor}, and integrating between $10^7$~Hz and $10^{11}$~Hz.  From this we find an efficiency of $\epsilon=0.03$.  Typical values of $\epsilon$ fall between 0.001 and 0.1~\citep{birzan}.  At the reference frequency, the core of the AGN is not visible and we can set $L_{rad,l}=L_{rad}$.  From the density profile we find $\rho_{ICM}=6.\times10^{-26}$~g~cm$^{-3}$.  We measure $r_c = 17$~kpc and $r_r = 7$~kpc.  The only unknown is $v_r$---the velocity of the plasma in the radio lobe.  In well-collimated jets, flow velocities are estimated to be between $0.3c$ and $0.7c$~\citep{jetha2006}.  In the lobes, these velocities are less restricted, but may be between $0.05c$ and $0.2c$~\citep{douglass}.  Using those values as the lower and upper limits, we find that the relative velocity between the galaxy and the ICM (equation~\ref{velocity}) is between 150~km~s$^{-1}$ and 300~km~s$^{-1}$, which is consistent with sloshing velocities~\citep{mendygral2012}.  As mentioned previously, the lobe is being displaced along the direction of gas flow, so it is likely that the sloshing in A2029 is distorting the radio lobe.

\section{Cooling vs. Heating in A2029}

The gas cooling time can be calculated from the following equation~\citep{sarazin}:
\begin{equation}
t_{cool}=8.5\times10^{10} \mathrm{~yr}\left(\frac{n_p}{10^{-3}\mathrm{~cm^{-3}}}\right)^{-1}\left(\frac{T_g}{10^8\mathrm{~K}}\right)^{1/2},
\label{cooling}
\end{equation}
where $n_p=1.21n_e$ is the proton number density and $T_g$ is the temperature of the gas.  From the temperature and density profiles we find that $T_g=6.1$~keV and $n_e=0.043$~cm$^{-3}$ at a radius of 13.8"~(20~kpc).  Inside this radius we calculate a cooling time of $\sim1.4\times10^9$~yr, which is much shorter than the Hubble time. This is another indicator (along with spectrally-determined the high mass deposition rate) that cooling is important in the cluster.  We can also calculate a mass deposition rate based on the X-ray luminosity of the gas.  For steady-state, isobaric cooling,  
\begin{equation}
L_{cool}=\frac{5}{2}\frac{kT}{\mu m_p}\dot{M}.
\label{rate}
\end{equation}
As above, $\mu$ is the mean atomic mass of the cluster (which we take to be $\mu$=0.6).  Inside this radius, $L_{cool}=1.4\times10^{45}$~erg~s$^{-1}$ (derived from the single APEC X-ray spectral fit).  Setting $kT=8.5$~keV (the temperature at $r=116"$), we find \textit{\.{M}}$=805$~$M_{\sun}$~yr$^{-1}$.  

As discussed in \S4.1, our spectroscopically derived cooling flow rate differs significantly from the value derived in \citet{a2029}.  We have performed a series of tests to determine the source of this discrepancy.  One possibility is simply that the differences are due to using a deeper set of observations here, as well as updated calibrations.  To test this we performed the spectral fits just on the 20 ks dataset, and found that mass deposition rate was \textit{\.{M}}$=378^{+313}_{-348}$~$M_{\sun}$~yr$^{-1}$, which, because of the large error bars, is consistent with \textit{\.{M}}$=56$~$M_{\sun}$~yr$^{-1}$. A second possibility is that there is another minimum in the parameter space that gives the lower cooling flow rate.  To test this we used both data sets.  Since \citet{a2029} uses the energy range 0.7-8~keV to do their spectral fits, we first tested the fit by using our data and fitting in that energy range.  Again, a high mass deposition rate was favored.  We also fixed the mass deposition rate to \textit{\.{M}}$=56$~$M_{\sun}$~yr$^{-1}$, then compared that fit with our fit.  Comparing with an F-test, the higher mass deposition rate is a better fit, with an F-test probability of 0.03\%.  We then looked at what happened when we fixed the upper energy to 7 keV, and varied the lower limit between 0.4 and 0.8~keV.  Again, the larger mass deposition rate was favored.  Lastly, we fit the spectrum within a smaller radius.  With a smaller radius we expect the mass deposition rate to be lower.  We find that it is; however, it is still consistent with a mass deposition rate in the hundreds of solar masses per year, with no overlap in the error bars with the lower rate.  We again fixed the mass deposition rate to the lower value, and still find that the higher value is a better fit.  We conclude that the high mass deposition rate is the better fit, although it is not very well constrained.  It is also likely that heating from the AGN is insufficient to offset the cooling.  Sloshing may supply an important, additional heating mechanism.

Despite the high mass deposition rate, there is very little star formation taking place in A2029.  \citet{hicks10} found a star formation rate of \textit{\.{M}}$=0.03-0.06$~$M_{\sun}$~yr$^{-1}$ using \textit{GALEX}.  There is also no evidence of $H\alpha$ emission from the central cD~\citep{mcdonald2010}.  One possible explanation for the apparent lack of star formation is heating by the central AGN.  As above, we estimate the energy injection from the AGN as $E=4PV$ and average over a typical source repetition rate of $5\times10^7$~yrs.  This yields $1\times10^{44}$~erg~s$^{-1}$, which is an order of magnitude lower than necessary to offset the cooling.  Thus A2029 provides a good example of a cluster where sloshing may be a significant component of heating that prevents large amounts of cooling~\citep{zuhone}.  We can contrast A2029 with the cluster SPT-CLJ2344-4243 ($z=0.596$), which has a mass deposition rate of  \textit{\.{M}}$=2700\pm700$~$M_{\sun}$~yr$^{-1}$, but also shows a large star formation rate of \textit{\.{M}}$=798\pm42$~$M_{\sun}$~yr$^{-1}$~\citep{phoenix1}.  The mass deposition rate was determined using the X-ray luminosity~\citep{phoenix}, rather than spectroscopically.  The cooling flow there is likely inducing the starburst, indicating that whatever feedback mechanism that is prevalent in low-redshift clusters such as A2029 is not present in the higher redshift system. 

\section{Discussion and Conclusions}

We have presented deep \textit{Chandra} observations of the cluster A2029.  We examine in detail the spiral structure that is seen in the X-ray image after beta-model subtraction.  This spiral is likely the result of an off-axis merger with a subcluster, as similar features are seen in simulations of such events.  The presence of the spiral highlights the fact that even in very relaxed clusters sloshing spirals are possible (A2029 is considered one of the most relaxed clusters in the Universe~\citep{buote}, and there is no optical evidence of substructure~\citep{dressler}).  The brightest part of the spiral (shown in the black contours in Figures~\ref{tmap},~\ref{abun}, and~\ref{entropy}) extends approximately 300~kpc from the cluster core to the brightest edge directly south.  Measured to the southeast edge, the full extent of the spiral reaches 400~kpc (green contours in Figures~\ref{tmap},~\ref{abun}, and~\ref{entropy}), making it the largest continuous sloshing spiral identified to date.  There is some evidence that the spiral in Perseus extends out to 700~kpc in the east and over a~Mpc in the west~\citep{simionescu}; however it is not continuous.  This may indicate that it is observed at a less face-on orientation than the spiral in A2029.   

The total cluster emission is best fitted by either an APEC+APEC model or an MKCFLOW+APEC model with absorption allowed to vary.  The APEC+APEC model with absorption fixed gives $kT_{low}=5.97^{+0.56}_{-0.68}$~keV and $kT_{high}=11.16^{+33.7}_{-1.63}$~keV, with normalization values that are within a factor of two.  The abundance for this model is $0.52^{+0.03}_{-0.02}$~$Z_{\sun}$.  The cooling flow model yields a high temperature component of $kT_{high}=10.08^{+2.30}_{-1.12}$~keV cooling to a temperature of $kT_{low}=4.24^{+0.80}_{-0.67}$~keV with a mass deposition rate of \textit{\.{M}}$=783^{+371}_{-268}$~$M_{\sun}$~yr$^{-1}$.  The abundance is $0.52\pm0.03$~$Z_{\sun}$.  While fairly high, this mass deposition rate is consistent with other studies of this cluster, particularly the ROSAT value of $556^{+215}_{-93}$~$M_{\sun}$~yr$^{-1}$~\citep{peres}. The ratio  $kT_{high}/kT_{low}\sim2.4$ is consistent with previous values of other clusters~\citep{peterson}.  The mass deposition rate found here differs significantly from the rate of 56~$M_{\sun}$~yr$^{-1}$ and the ratio $kT_{high}/kT_{low}>50$ found by~\citet{a2029}.  A series of tests shows that the high mass deposition rate of \textit{\.{M}}$=783^{+371}_{-268}$~$M_{\sun}$~yr$^{-1}$ is a better fit to the data. 

We also find that the center of the cluster is most likely dominated by the thermal component of the gas and do not find evidence for non-thermal emission from the AGN. 

The surface brightness profile shows a clear brightness excess in the regions where the spiral is seen.  We also see a decrease in temperature in the area of the spiral compared to the average cluster temperature, and an increase in density.  Both of these are consistent with simulations of sloshing.  These features are most obvious in the N region of the cluster, where the brightness excess is the largest.  The pressure varies smoothly across the spiral, consistent with a cold front.

These features are also seen in the spectral maps.  The spiral can be seen tracing areas of cooler temperature, higher abundance, and lower entropy, which is consistent with a scenario where cool, metal-rich, central gas is displaced to larger radii.  Temperature maps from \textit{XMM-Newton} also show areas of low temperature gas tracing a spiral structure to the south, however there is no explicit mention of sloshing~\citep{xmm}.

The simulations performed by \citet{am} and \citet{zuhone} start with initial conditions for their main clusters that closely match the conditions in A2029.  Thus, we can compare our observations to their simulations to put rough constraints on the merger history of A2029.  The simulations that appear most similar to our observations have a mass ratio of 5 and an impact parameter of 500~kpc.  From these simulations, it appears that the merger in A2029 occurred between 2-3~Gyr ago.  To constrain the geometry of the merger, we compare our observations to A496~\citep{a496}, which is an elliptical cluster similar to A2029.  The ellipticity introduces uncertainty in the determination of the orbit orientation, however \citet{a496} find that the large-scale structure of their residuals favors the fiducial case, where the subcluster crosses along a diagonal orbit from the SW to the N/NE.  This is what we compare with A2029.  Doing this, we conclude that the subcluster most likely followed an orbital path from the southeast to the northwest, passing west of the cluster center.

Sloshing features are thought to be associated with radio mini-halos, with these halos tracing the properties of the spiral~\citep{mazzotta}. A2029 is host to such a halo, and the morphology of the halo is similar to the morphology of the spiral~\citep{halo}.  The connection between the mini-halo and the spiral is not well understood, and more observations in the radio are needed to fully explore this phenomenon.  It is possible that some of the energy from the sloshing is going into re-accelerating the radio-emitting relativistic particles.

The sloshing spiral is most likely distorting the central AGN, giving it a WAT morphology.  This distortion will be useful for finding clusters at high redshift.  Bent, double-lobed radio sources (such as the one at the center of A2029) are found in clusters $\sim70\%$ of the time, and thus are good tracers of clusters over a wide range of redshifts~\citep{wing2011}.  The~\citet{mendygral2012} simulations show that these bent AGN can be found in relatively relaxed clusters, as well as in clusters where mergers have recently occurred.  This will allow us to probe clusters with varying properties and to explore their evolution with redshift.
\\
\\
RPM and ELB were partially supported by NASA through the Astrophysics Data Analysis Program, grant number NNX10AC98G, and through NASA award RSA No. 1440385 issued by JPL/Caltech.  SR is supported by the Chandra X-ray Center through NASA contract NAS8-03060 and the Smithsonian Institution, and in part by NASA grants GO1-12104X and NNX10AR29G.  TEC was partially supported by NASA through Chandra Grant GO4-5149X issued by the Chandra X-ray Observatory Center for and on behalf of NASA under contract NAS8-39073. Basic research in radio astronomy at the Naval Research Laboratory is funded by 6.1 Base funding.  

Funding for SDSS-III has been provided by the Alfred P. Sloan Foundation, the Participating Institutions, the National Science Foundation, and the U.S. Department of Energy Office of Science. The SDSS-III web site is http://www.sdss3.org/.

SDSS-III is managed by the Astrophysical Research Consortium for the Participating Institutions of the SDSS-III Collaboration including the University of Arizona, the Brazilian Participation Group, Brookhaven National Laboratory, University of Cambridge, Carnegie Mellon University, University of Florida, the French Participation Group, the German Participation Group, Harvard University, the Instituto de Astrofisica de Canarias, the Michigan State/Notre Dame/JINA Participation Group, Johns Hopkins University, Lawrence Berkeley National Laboratory, Max Planck Institute for Astrophysics, Max Planck Institute for Extraterrestrial Physics, New Mexico State University, New York University, Ohio State University, Pennsylvania State University, University of Portsmouth, Princeton University, the Spanish Participation Group, University of Tokyo, University of Utah, Vanderbilt University, University of Virginia, University of Washington, and Yale University.

\bibliographystyle{apj}
\bibliography{a2029}

\end{document}